\documentclass[conference]{IEEEtran}

\IEEEoverridecommandlockouts
\usepackage{cite}
\usepackage{amsmath,amssymb,amsfonts}

\usepackage{algorithm}

\usepackage{algorithmicx}
\usepackage{algpseudocode}
\usepackage{graphicx}
\usepackage{tikz}
\usepackage{textcomp}
\usepackage{xcolor}

\usepackage[top=0.75in, bottom=0.98in, left=0.57in, right=0.57in]{geometry}

\usepackage{lipsum}    

\begin{document}

\IEEEpubid{\makebox[\columnwidth]{%
   \footnotesize \copyright 2025 IEEE . Personal use is permitted. Permission required for reuse.}%
 \hspace{\columnsep}\makebox[\columnwidth]{}
}

\title{CNN-Enabled Scheduling for Probabilistic Real-Time Guarantees in Industrial URLLC\\
{\footnotesize \textsuperscript{}}
}
\author{
   \IEEEauthorblockN{Eman Alqudah, Ashfaq Khokhar}
    \IEEEauthorblockA{
       \textit{Department of Electrical and Computer Engineering}\\
        \textit{Iowa State University, USA}\\
        \{alqudah, ashfaq\}@iastate.edu}
}        

\maketitle

\begin{abstract}
Ensuring packet-level communication quality is vital for ultra-reliable, low-latency communications (URLLC) in large-scale industrial wireless networks. We enhance the Local Deadline Partition (LDP) algorithm by introducing a CNN-based dynamic priority prediction mechanism for improved interference coordination in multi-cell, multi-channel networks. Unlike LDP's static priorities, our approach uses a Convolutional Neural Network and graph coloring to adaptively assign link priorities based on real-time traffic, transmission opportunities, and network conditions. Assuming that first training phase is performed offline, our approach introduced minimal overhead, while enabling more efficient resource allocation, boosting network capacity, SINR, and schedulability. Simulation results show SINR gains of up to 113\%, 94\%, and 49\% over LDP across three network configurations, highlighting its effectiveness for complex URLLC scenarios.
\end{abstract}

\section{Introduction} Industrial ultra-reliable, low-latency communications (URLLC), driven by 5G and future advancements, are expected to significantly enhance industrial cyber-physical systems' performance, adaptability, and resilience. These systems must meet strict timing requirements in critical applications such as sensing, robotic control, process automation, and power grid management \cite{b1}. In Extended Reality (XR) applications \cite{b2}, timely packet delivery is essential for smooth 3D scene reconstruction, as delays or packet losses can degrade user experience \cite{b2}. Similarly, packet loss can compromise system stability and safety in networked industrial control. Achieving real-time communication in multi-cell industrial wireless networks is challenging due to transmission delays and the time required for data processing and transmission \cite{b3}.

\subsection{Related Work} Research on real-time communication guarantees in wireless networks has been explored through optimization~\cite{b4,b5} and machine formulations~\cite{b6,b7}. These formulations have been applied to resource allocation and scheduling problems, such as earliest-deadline-first (EDF) and rate-monotonic (RM) \cite{b3,b8}. Concurrently, studies have also focused on long-term real-time guarantees, like mean delay and age-of-information (AoI) \cite{b9,b10}. Despite significant contributions, limitations remain, including challenges in ensuring reliability and latency, especially in multi-cell environments. Some recent work \cite{b11,b12} has examined 5G configured grant (CG) scheduling for real-time guarantees, but CG scheduling applies only to 5G uplink transmissions and lacks flexibility.
One of the recent work~\cite{b13} based on Local Deadline Partitioning (LDP) has provided significant insights into optimizing task scheduling in multi-cell systems. Their approach focuses on efficiently partitioning communication tasks to meet strict deadlines, particularly in systems with varying resource availability. The Local Deadline Partitioning (LDP) algorithm\cite{b13} has shown effectiveness in real-time scheduling However, several limitations restrict its applicability in dynamic multi-cell environments. First, LDP relies on static priorities assigned to communication links, making it less adaptable to fluctuating traffic conditions and varying resource availability. In addition, its computational complexity increases with the number of tasks and cells, posing scalability challenges. The method also depends on predefined heuristics, which may lead to suboptimal decisions, especially in unpredictable network conditions.
Deep learning techniques, particularly Convolutional Neural Networks (CNNs) are well-suited for resource allocation in wireless networks due to their ability to learn high-dimensional spatial patterns and dependencies. In this paper, we develop a significant enhancement of the LDP approach using CNNs based predictive models to dynamically determine link priorities in allocating resources and providing real-time guarantees. In addition, we employ graph-coloring technique to minimize interference among competing nodes.
By training on historical data and continuously updating with new traffic and interference patterns, our work introduces a novel scheduling mechanism that optimally partitions deadlines while addressing interference and traffic fluctuations, making it a robust solution for industrial URLLC scenarios.
The rest of the paper is organized as follows. Section II presents the system model and the problem statement. Section III introduces the proposed solution, which incorporates the CNN-Based Modified Local-Deadline-Partition (LDP) Algorithm and CNN models architecture. Section IV details the experimental study, while Section V discusses the results. Finally, the conclusion offered in Section VI.

\section{System Model and Problem Statement}
In order to properly analyze and solve the scheduling problem, we present a detailed network model that captures the key characteristics and dynamics of the communication environment, including the relationship among interferences across different network links.

The network comprises $m$ base stations (BSes) and $n$ user equipment (UEs). Communication between BSes and UEs is realized using cellular links, while direct communication between UEs is accomplished using device-to-device (D2D) links. This wireless network can be represented as a graph $G = (V, E)$, where $V$ denotes the set of nodes (including both BSes and UEs), and $E$ represents the set of wireless links.
The edge set $E$ consists of edges between node pairs that are within each other's communication range. The network operates over $N$ non-overlapping frequency channels, referred to as RBs. Time is slotted and synchronized across all transmitters and receivers, ensuring that wireless transmissions are scheduled in both the frequency and time domains. Each transmission is assigned a specific frequency channel and time slot, with all time slots having uniform duration. Within a single time slot, a transmitter can complete the transmission of one packet.

To achieve predictable communication, this paper employs  the same interference model as used in \cite{b13} to determine the conflict set for each link. Specifically, the conflict graph \( G_c = (V_c, E_c) \) is constructed for the network \( G \), where each node in \( V_c \) represents a unique communication link in \( G \). An edge \( (i, j) \in E_c \) exists if links \( i \) and \( j \) interfere with each other—that is, if the transmitter of link \( i \) (or link \( j \)) falls within the exclusion region of link \( j \) (or link \( i \)).  

For a given link \( i \), let \( M_i \) represent the set of links that interfere with it, defined as:

\[
M_i = \{ j : (i, j) \in E_c \}
\]

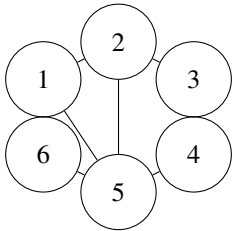
\begin{figure}[h]
    
\begin{center}
\begin{tikzpicture}[scale=0.5, every node/.style={circle, draw, minimum size=1cm}]

\node (1) at (0,2) {1};
\node (2) at (2,3) {2};
\node (3) at (4,2) {3};
\node (4) at (4,0) {4};
\node (5) at (2,-1) {5};
\node (6) at (0,0) {6};

\draw (1) -- (2);
\draw (2) -- (3);
\draw (2) -- (5);
\draw (3) -- (4);
\draw (4) -- (5);
\draw (5) -- (6);
\draw (6) -- (1);
\draw (1) -- (5); 

\end{tikzpicture}
\end{center}
    \caption{Conflict graph with 6 nodes, where each node represents a communication link, and edges indicate interference relationships.}
    \label{figure:conflictgraph}
\end{figure}

As an illustration, consider the conflict graph with 6 nodes shown in Figure~\ref{figure:conflictgraph}, where each node corresponds to a link in network \( G \). The interference set for link 2 is:
\[
M_2 = \{1, 3, 5\}
\]
This means that if link 2 is active, links 1, 3, and 5 cannot be active simultaneously in the same time slot and frequency channel.

Similar to~\cite{b13}, in our work we use the PRK (Physical-Ratio-K) interference model~\cite{b13} to ensure predictable communication reliability. 
Given a network \( G = (V, E) \), where each link \( i \) has a traffic demand \( d'_i \), a local deadline \( D_i \), and a priority class \( p_i \), the scheduling problem aims to allocate the available resource blocks (RBs) to the communication links while satisfying the real-time constraints and minimizing interference. The main challenge is to develop an efficient algorithm that assigns RBs to links in a way that meets their traffic demands within their deadlines, accounts for interference between links, and prioritizes high-demand links based on their priority classes. Specifically, the algorithm must ensure that the effective demand of each link, which is influenced by both the traffic demand and the signal-to-noise ratio (SNR), does not exceed the available RBs, while adhering to the local deadlines. The scheduling problem involves balancing the allocation of limited resources to maximize the system's overall performance, ensuring predictable communication reliability, and meeting all the real-time constraints imposed by the network.

The goal is to find the set of RB allocations \( \{ r_i \} \) for all communication links \( i \in V \) that minimizes interference while satisfying the above constraints. The objective is to ensure efficient resource utilization by allocating resource blocks in such a way that interference between links is minimized, traffic demands are met, and deadlines for all communication links are respected. Mathematically, this optimization problem can be formulated as:  

\( \min_{r_i} \sum_{i \in V} \sum_{j \in V, j \neq i} I(r_i, r_j) \)

\section{Proposed Solution}
In this section, we present our proposed approach for efficient resource allocation and scheduling in communication networks. 

We introduce the use of a Convolutional Neural Network (CNN) and graph-colorig based architecture to predict link priorities and share transmission states among neighboring links. 
\subsection{CNN Architecture for Priority Prediction and Resource Block Selection Models}

We use two different convolutional neural network (CNN) models, one each for priority prediction and resource block (RB) selection in real-time resource allocation. 
The priority prediction model provides priority values for each link at each time step, facilitating an informed scheduling mechanism. The RB selection model predicts optimal allocation of resource blocks, ensuring efficient spectrum utilization and minimizing interference. Together, these models contribute to improving scheduling efficiency and real-time decision making in dynamic wireless communication environments.

The CNN architecture is chosen for its capability to capture spatial relationships within structured input data, making it well-suited for analyzing link-specific features in wireless communication systems. Both models process input data as 4D tensors, where each sample consists of multiple links and their associated features, reshaped to include a channel dimension to enable convolutional operations. These models are trained on preprocessed dataset and employ a custom loss function that balances categorical cross-entropy with an L1 penalty. 

The prediction models for Link Priority and RB Selection  are designed to estimate priority values based on traffic demand, remaining time, and other link-specific attributes. As you can see in Figure 2, the model begins with a convolutional layer containing 32 filters of size \(3 \times 3\) with ReLU activation and the same padding to preserve the spatial dimensions. This is followed by a max pooling layer with a pool size of \(2 \times 2\) and strides of (1,1) to down-sample the feature maps while maintaining structural consistency. 

A second convolutional layer with 64 filters is applied, again with \(3 \times 3\) kernels and ReLU activation, followed by another max-pooling layer with identical parameters. The extracted features are then flattened and passed through two fully connected layers with 256 and 128 neurons, respectively, using ReLU activation. The final output layer employs a softmax activation function to predict priority values across multiple categories, with its output reshaped to match the target priority matrix. The two models are trained with different data sets and individually tailored for each use case. Algorithm 1 shows the details of the CNN-Based LDP Algorithm.

\begin{algorithm}
\caption{CNN Based LDP Algorithm}
\begin{algorithmic}[1]
\State \textbf{Input:} $A_{i,1}$ (arrival time of first packet for link $i$), $M_i$ (set of interfering links of $i \in E$), $T_l, D_l$ (period and relative deadline for $l \in M_i \cup \{i\}$), $X_{i,t}$ (local traffic demand at $i$), $\text{State}[l][\text{rb}][t]$ (transmission state for $l \in M_i \cup \{i\}$), $\text{Prio}[l][t]$ (priority for $l \in M_i \cup \{i\}$).
\State \textbf{Output:} Transmitter and receiver actions for link $i$ at time $t$.
\State $\text{State}[i][\text{rb}][t] = \text{UNDECIDED}, \forall \text{rb} \in \text{RB}$
\State $\text{Prio}[i][t] = \text{CNN-Predict}(\{X_{i,t}, d''_{i,t} - t, \text{State}[l][\text{rb}][t], \text{Prio}[l][t]\}_{l \in M_i})$
\State Share $\text{Prio}[i][t]$ with links in $M_i$
\State $\text{done} = \text{false}$
\While{$\text{done} == \text{false}$}
    \State $\text{done} = \text{true}$
    \For{each $\text{rb} \in \text{RB}$ in increasing order of $\text{ID}(\text{rb})$}
        \If{Updates received for $\text{State}[l][\text{rb}][t]$ or $\text{Prio}[l][t]$ from $l \in M_i$}
            \State Update local copy of $\text{State}[l][\text{rb}][t]$ or $\text{Prio}[l][t]$
        \EndIf
        \If{$X_{i,t} == 0$ and $\text{State}[i][\text{rb}][t] == \text{UNDECIDED}$}
            \State $\text{State}[i][\text{rb}][t] = \text{INACTIVE}$; \textbf{break}
        \EndIf
        \If{$\exists l \in M_i : \text{State}[l][\text{rb}][t] == \text{ACTIVE}$}
            \State $\text{State}[i][\text{rb}][t] = \text{INACTIVE}$; \textbf{break}
        \EndIf
        \If{$\text{State}[i][\text{rb}][t] == \text{UNDECIDED}$ and $\text{CNN-Rank}(\text{Prio}[i][t], \text{Prio}[l][t])$ holds for all $\text{UNDECIDED}\ l \in M_i$}
            \State $\text{State}[i][\text{rb}][t] = \text{ACTIVE}$; $X_{i,t} = X_{i,t} - 1$
        \EndIf
        \If{$\text{State}[i][\text{rb}][t] == \text{UNDECIDED}$}
            \State $\text{done} = \text{false}$
        \EndIf
    \EndFor
    \State Use $\text{CNN-Predict}$ to estimate likelihood of $\text{State}$ changes
    \If{$\text{Predicted Changes are Significant}$}
        \State Share $\text{State}[i][\text{rb}][t], \forall \text{rb} \in \text{RB}$ with $M_i$
    \Else
        \State Skip sharing updates for $\text{State}[i][\text{rb}][t]$
    \EndIf
\EndWhile
\end{algorithmic}
\end{algorithm}

After each round of resource block allocation, the state of the resource block (whether it is ``ACTIVE'' or ``INACTIVE'') is shared with the neighboring links in $M_i$. If the CNN model predicts significant changes to the transmission state, these updates are promptly shared. If not, the sharing step is skipped to save network resources.

The process continues iteratively until all links have been assigned resource blocks or marked as inactive. The algorithm ensures that all deadlines are met while minimizing interference between links and optimizing resource block usage.




\subsubsection{Optimization and Loss Functions}
For training, we employ \textit{categorical cross-entropy loss}, which is well-suited for multi-class classification tasks. The RB selection model also incorporates a \textit{custom LDP-shared loss function}, which combines categorical cross-entropy with a penalty term to enhance resource-sharing stability based on local deadline partitioning (LDP). The models are optimized using stochastic gradient descent (SGD), ensuring efficient convergence while maintaining computational feasibility.

\subsubsection{Training Dataset}
The dataset used in this experiment consists of 1000 samples with a fixed number of links. The set of features is represented by various parameters, including work densities (\( X \)), priority values (\( y_{\text{priority}} \)), interference levels (\( M_i \)), traffic demands (\( d' \)). Based on these demands, local deadlines are generated by scaling the traffic demand values with a random factor between 1.2 and 2.0, ensuring variability in deadline constraints. The link positions are also extracted to model interference relationships among links. This dataset is used to train and validate a convolutional neural network (CNN) model. The model predicts Resource Block (RB) priorities based on traffic demand, interference levels, and deadlines. These predictions used for RBs selection to minimize interference while ensuring that all links meet their deadlines, optimizing\newpage 
resource allocation and maintaining system stability.


\begin{figure}[!htbp]

    \centering
    \includegraphics[width=\columnwidth ,height=0.25\textheight]{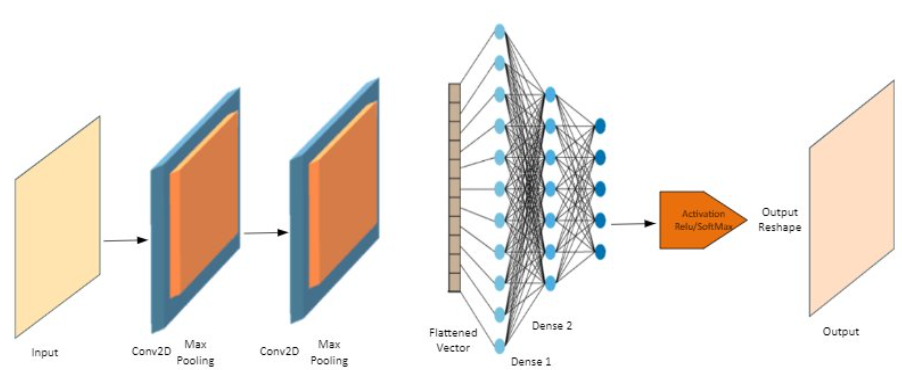} 
    
    \caption{Architecture of CNN Models.}
     \label{fig:model}
\end{figure}

\vspace{-10pt}

\subsection{Graph-Coloring Based Resource Allocation with CNN Prediction}
Graph coloring is an effective method for modeling interference constraints in resource allocation~\cite{b14}. In this approach, links are represented as nodes and interference relationships are modeled as edges between nodes. By ensuring that no two interfering links share the same resource block (RB), graph coloring naturally aligns with the need to prevent conflicting  
links from using the same RB simultaneously.

Integrating graph coloring with resource allocation offers several key benefits. It effectively models interference constraints by ensuring that interfering links (connected nodes) are assigned different RBs, a requirement in the allocation policy. The approach also enables priority-driven scheduling, where higher priority links, predicted by a CNN are allocated RBs first, with lower priority links adjusted to avoid interference and conflict. Furthermore, graph coloring efficiently resolves conflicts by finding alternative RBs when a predicted RB conflicts with a neighbor’s allocation, ensuring that all resource constraints are satisfied.
The interference graph is constructed, where each node represents a communication link, and edges are added between nodes that interfere with each other. This graph serves as the basis for resource block (RB) allocation.

The nodes are processed in descending order of priority, and the CNN model provides an initial RB prediction. For each node (link), the predicted RB is used as the first choice. If the predicted RB conflicts with neighboring nodes, alternative RBs are considered. If no RB satisfies the required constraints, such as interference, capacity, and deadlines, the node is marked as "unallocated."

\section{Experimental Evaluation Setup}
In this section, we evaluate the proposed scheduling technique that integrates convolutional neural networks (CNN) and graph coloring techniques to optimize resource allocation in multi-cell industrial wireless networks. 


We consider three networks of different sizes to represent various real-time network scenarios. The network size, number of channels, link/node spatial distribution density, and number of conflicting links per link are chosen based on industrial URLLC settings similar to the setting in \cite{b13}.

    \textbf{Network 1}: 91 wireless nodes are deployed in a $120 \times 120$ square meter region, generating 83 links. The network is organized into nine cells (3×3 grid), each with a base station (BS).
    \textbf{Network 2}: 151 wireless nodes are deployed in a $120 \times 120$ \newpage square meter region, generating 163 links. The same 3×3 cell grid structure applies.
\textbf{Network 3}: 320 wireless nodes are deployed in a $240 \times 240$ square meter region, generating 324 links. The network consists of 36 cells (6×6 grid).

We apply the Wireless Industrial Indoor path loss model to determine the interference effect among links. Regarding channel allocation, we assume a 5G Numerology 4 setting where each resource block (RB) occupies 2.8 MHz. With a total of 20 MHz bandwidth, there are 7 RBs available ($N=7$). To simulate various industrial URLLC scenarios, the available channels range from 3 to 11.
%

\subsection{System Evaluation and Performance Metrics}
To assess the effectiveness of the proposed scheduling algorithm, we evaluate key performance metrics that reflect network efficiency, reliability, and schedulability. In addition, we also assess the quality of the links using SINR.


We incorporate traffic demand as a key criterion for determining a link’s schedulability. A link is considered schedulable if its traffic demand can be accommodated within its allocated capacity before the deadline. Mathematically, for a given link \( i \), if the ratio of its traffic demand \( X_i \) to its link capacity \( C_i \) is less than or equal to its relative deadline \( D_i \), then the link can be scheduled within its deadline, i.e., 
%
\( \frac{X_i}{C_i} \leq D_i \)
%
The \textit{schedulable ratio} quantifies the number of successfully scheduled links without violating deadlines, and it is defined as follows:
\[
\text{Schedulable Ratio} = \dfrac{\sum_{i=1}^{N} \mathbb{1} \left( \frac{X_i}{C_i} \leq D_i \right)}{N}
\]


%
%

To assess Signal-to-Interference-plus-Noise Ratio (SINR), the quality of received signals, we measure SINR as follows :
$ \text{SINR} = \frac{P_{signal}}{P_{interference} + P_{noise}}$
where $P_{signal}$ is the received signal power, $P_{interference}$ is the sum of interference power from other links that share the same resource block, and $P_{noise}$ represents the background noise power. 

The received signal power for the \( i \)th link, denoted as \( P_{\text{signal},i} \), is determined by the transmitted power \( P_{\text{tx}} \), the distance \( d_i \) between the transmitter and receiver of the \( i \)th link, and the path loss exponent \( \alpha \). It is expressed as:
%
$    P_{\text{signal},i} = \frac{P_{\text{tx}}}{d_i^\alpha}$
%
The transmitted power \( P_{\text{tx}} \) is considered constant for all nodes and the  path loss exponent \( \alpha \) characterizes the rate at which the signal attenuates with distance, typically ranging from 2 in free-space environments to values above 3 in obstructed or indoor scenarios. As the distance \( d_i \) increases, the received signal power decreases exponentially due to propagation losses.

The interference power experienced by the \( i \)th link, denoted as \( P_{\text{interference},i} \), is caused by other links assigned to the same resource block and that interfere with link \( i \) based on the interference graph \( G \). The set of interfering links for \( i \) is denoted as \( \mathcal{I}_i \), which consists of all links that share an edge with \( i \) in the interference graph. This formulation captures the impact of co-channel interference while respecting the interference graph structure. The interference power is thus given by: 
\[
P_{\text{interference},i} = \sum_{\substack{j \in \mathcal{I}_i \\ \text{RB}_i = \text{RB}_j}} \frac{P_{\text{tx}}}{d_{i,j}^\alpha}
\]

\newpage
The noise power \( P_{\text{noise}} \) in a communication system is the power associated with the thermal noise in the system. It can be calculated using the following formula:
 \(   P_{\text{noise}} = k T B \)
where:
- \( k \) is Boltzmann's constant, \( k = 1.38 \times 10^{-23} \, \text{J/K} \),
- \( T \) is the system temperature in Kelvin. A common value for communication networks is \( T = 290 \, \text{K} \), which is considered room temperature,
- \( B \) is the bandwidth of the system in Hertz (Hz). We have used the same assumption as in~\cite{b13} and assumed bandwidth \( B = 20 \, \text{MHz} \).


\subsubsection{Link Reliability}
Link reliability is defined as the fraction of links that achieve SINR above a predefined threshold ($\gamma_{th}$). Given a set of SINR values, the reliability score is computed as:

\( R_{reliability} = \frac{\sum (\text{SINR} \geq \gamma_{th})}{N_{links}} \)
where $\gamma_{th} = 15$ dB. 

The estimated network capacity is derived by considering the number of schedulable and reliable links:
\( C_{network} = N_{links} \times R_{schedulable} \times R_{reliability} \)

This metric reflects the effective number of links capable of successful transmission in a given time slot.

The transmission latency for each link is estimated using the Shannon capacity formula:
\( C = B \log_2(1 + 10^{\text{SINR}/10}) \)
where $B$ is the bandwidth. The transmission time is then computed as:
\( L = \frac{S_{packet} \times 8}{C} \)
where $S_{packet}$ is the packet size in bytes. The BER for a modulation scheme (e.g., 16-QAM) is estimated using an error function approximation.

The Packet Error Rate (PER) is calculated as:
\( \text{PER} = 1 - (1 - \text{BER})^{N_{bits}} \)
where $N_{bits}$ is the packet size in bits. The estimated number of retransmissions is computed as:
\( R_{retrans} = \frac{1}{1 - \text{PER}} \)
A lower PER leads to fewer retransmissions, improving network efficiency.

\section{Results and Discussion}
This section presents the evaluation and analysis of the performance of our proposed CNN/graph-coloring-based scheduling solution. The results are based on simulations conducted under various network configurations, focusing on key performance metrics such as receiver-side SINR, prediction accuracy, and resource block selection efficiency. First, we outline the simulation setup used to generate the results, followed by a detailed presentation of the performance of the CNN pre-trained models across different experiments. The subsequent subsections discuss the observed trends, the significance of the improvements, and the implications of our findings.

\subsection{Simulation Setup}

The simulation experiments were conducted using Google Colab, a cloud-based platform that provides computational resources for machine learning and data processing tasks. The code was executed in Python, utilizing libraries such as TensorFlow, PyTorch, NumPy, and SciPy for efficient implementation. Hardware acceleration, including GPU support, was enabled when necessary to optimize performance.


\subsection{Performance of CNN pre-trained Models}
The training results indicate that our proposed models achieved exceptionally high accuracy, consistently reaching approximately 99.97\% across multiple epochs in both models. The loss values demonstrate a downward trend, showing continuous improvement in model performance. Initially, the training loss was relatively higher, but as the training progressed, it gradually decreased to approximately 0.0005 in later epochs. The validation loss followed a similar decreasing pattern, reaching values as low as $3.3563 \times 10^{-4}$.

\begin{figure}[!t]
    \centering
    \includegraphics[width=3.55in]{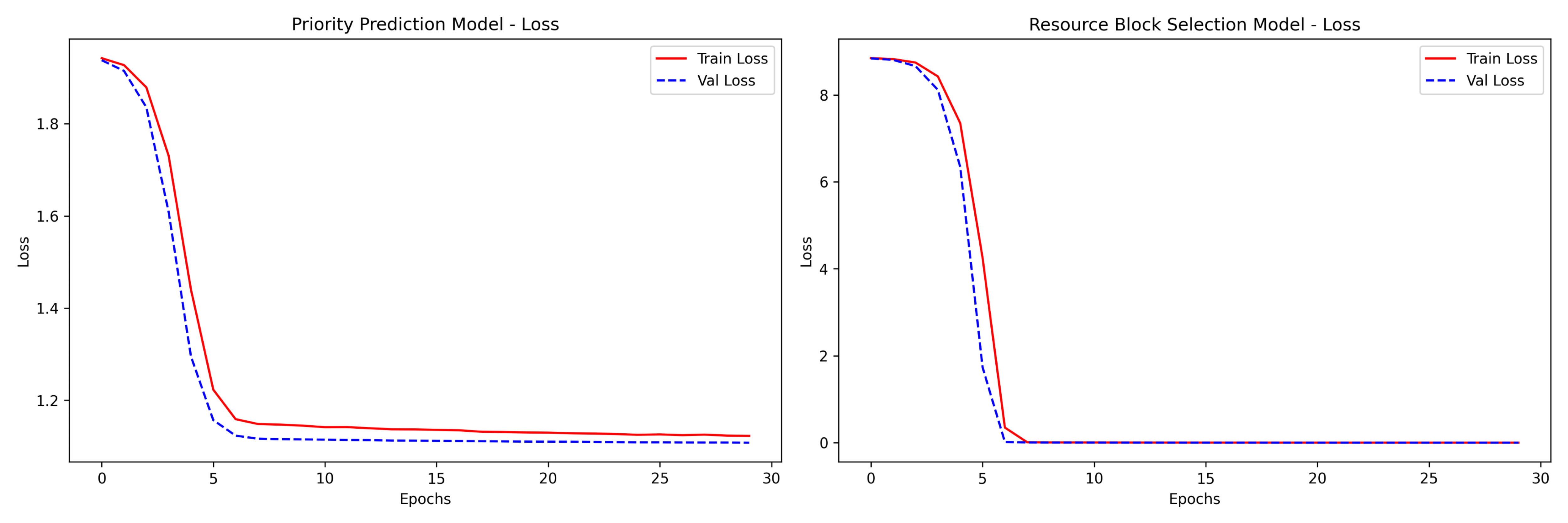} 
    \caption{Training Losses of our proposed CNN Models: priority prediction model and Resource Block Selection model.}
    \label{Lossmodel}
\end{figure}

Figure~\ref{Lossmodel} illustrates the training and validation loss curves for both the priority prediction model and the resource block selection model. The priority prediction model shows a sharp decline in the initial epochs, followed by a gradual stabilization at a low value, indicating effective learning. The close alignment between training and validation loss suggests minimal overfitting and good generalization. Similarly, the resource block selection model exhibits a decreasing trend of loss, with an initially rapid drop before reaching a stable value. The final loss values confirm that both models have successfully captured the underlying data patterns. 
The training phase of the resource block selection model, conducted offline on an Intel CPU E3-1240 v6 @ 3.70 GHz system with 16 GB of RAM, requires 306 seconds, whereas the inference phase, performed online, executes in just 0.47 seconds. The system operates on a 64-bit architecture without touch or pen input capabilities. This inference time can be further reduced significantly using dedicated hardware accelerators, making the model well-suited for real-time decision-making in dynamic wireless environments.

The steady reduction in loss and the stability of accuracy suggest that the chosen learning rate of $1.0 \times 10^{-4}$ was effective in optimizing the model. The minimal gap between training and validation loss further supports the claim that the model is well-fitted to the data and exhibits strong generalization capabilities.

\subsection{Receiver-Side SINR Performance}
To evaluate the impact of interference coordination on the performance of our proposed solution, we examined the receiver-side SINR performance and comparison with the LDP algorithm across three distinct network configurations. In Network 1, which consists of 83 links, we observed a significant improvement in the Signal-to-Noise Ratio (SINR). The mean SINR achieved by the CNN-based method was 32.09 dB, with the 25\%-75\% range spanning from 31.28 dB to 32.88 dB. In contrast, the mean SINR for the LDP algorithm in Network 1 was 15.09 dB, with the 25\%-75\% range between 14.13 dB and 15.95 dB. This comparison highlights the substantial gain in SINR achieved with our CNN-based method. For Network 2, consisting of 151 links, the mean SINR was 28.75 dB, with the 25\%-75\% range between 28.12 dB and 29.76 dB. LDP in this network showed a mean SINR of 14.79 dB, with the 25\%-75\% range from 13.94 dB to 16.12 dB. These results indicate that, even in larger networks, our CNN-based method outperforms the LDP algorithm, showcasing improved interference coordination. Finally, for Network 3, with 320 links, the mean SINR was 22.81 dB, and the 25\%-75\% range spanned from 22.31 dB to 22.70 dB. In contrast, the LDP method for Network 3 produced a mean SINR of 15.31 dB, with the 25\%-75\% range between 14.03 dB and 16.59 dB. Although the SINR values in our CNN-based solution are lower than those in Network 1, they are still significantly better than the LDP results across all three networks, underscoring the superior interference coordination of our approach.

\begin{figure}[!t]
\vspace{-5pt}
\centering
\includegraphics[width=3.55in]{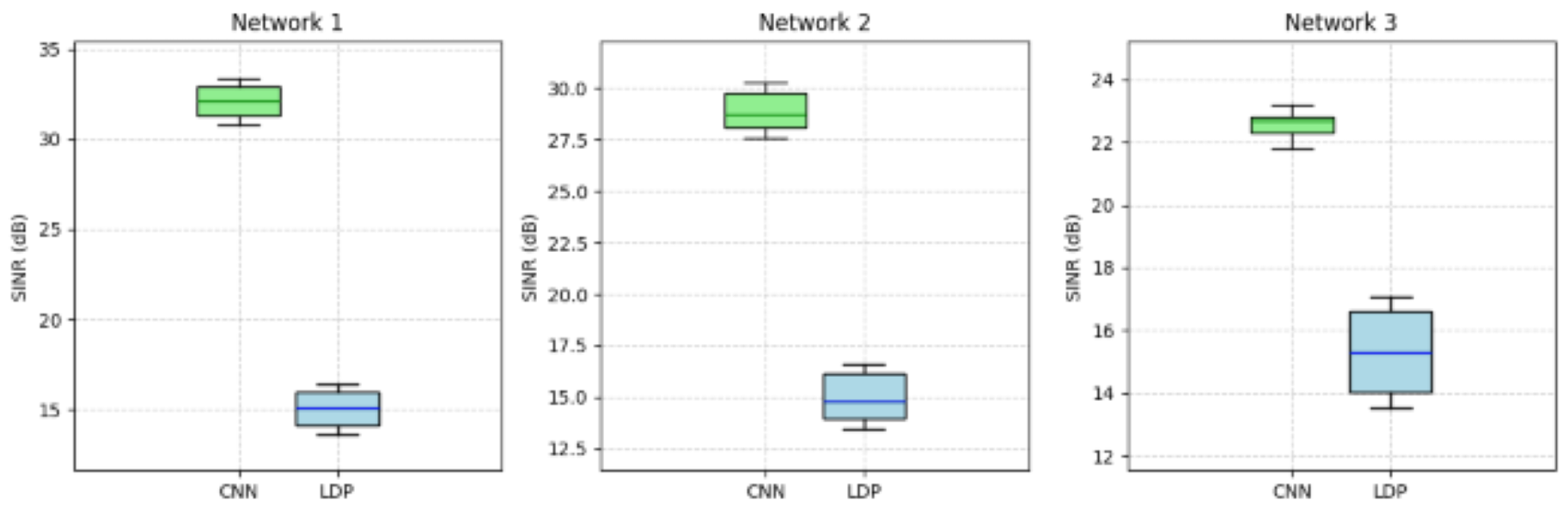}
\caption{Interference Effect on Receiver-Side SINR.}
\label{sinrcomparison}
\vspace{-10pt}
\end{figure}

Figure \ref{sinrcomparison} shows the examined  receiver-side SINR performance of our method in comparison to the LDP algorithm across three distinct network configurations. These results indicate a substantial improvement in SINR with our CNN-based approach and highlight the ability of our proposed solution to provide efficient interference coordination, improving the overall SINR and ensuring high network performance across varying network sizes.

We also assessed the impact of receiver-side SINR on real-time schedulability.
Our CNN-based algorithm demonstrated a consistent 100\% schedulability ratio across all network configurations and varying number of channels. This is a significant improvement over the existing state of the art. 

%

Moreover, in terms of reliability, the results from Networks 1, 2, and 3 revealed that all links had SINR values greater than 15 dB, resulting in a fraction of reliable links of 100\%, ensuring that each network consistently supports a large number of reliable communications. For Network 1, with 83 effective links, the average packet transmission latency was 0.04 ms, and the average bit error rate (BER) was 1.522233e-02. Similarly, for Network 2, the network capacity increased to 151 effective links, with the average latency and bit error rate remaining comparable at 0.04 ms and 2.124387e-02, respectively. Network 3, with 320 effective links, also demonstrated robust performance, with a slightly higher average latency of 0.05 ms and a higher bit error rate of 3.851555e-02. Despite these differences, all networks achieved the target performance, reinforcing the scalability and efficiency of our CNN-based solution across varying network sizes.

\section{Conclusion}

This paper introduced a machine learning-based solution, utilizing Convolutional Neural Networks (CNNs), for interference coordination in multi-cell, multi-channel networks supporting large-scale, industrial URLLC applications with diverse real-time requirements. The proposed CNN solution enhances interference mitigation, leading to significant improvements in SINR, real-time capacity, and network reliability. Our approach addresses the challenges of inter-cell interference and significantly improves network capacity, SINR, and schedulability. The numerical results show that our CNN-based method achieves up to 113\%, 94\%, and 49\% improvements in SINR over LDP in the three different network  configurations, respectively, highlighting its superiority in handling complex URLLC applications. Our results also demonstrate that this machine learning approach consistently outperforms traditional solutions, ensuring high network performance and schedulability. This highlights the transformative potential of machine learning in optimizing interference management, advancing the scalability and efficiency of complex communication networks.

\vspace{12pt}
\color{red}

\end{document}